\documentstyle[preprint,prl,aps]{revtex}

\begin{document}

\draft

\preprint{Submitted to Physical Review Letters}

\title{
The Role of the Environment in Chaotic Quantum Dynamics}

\author{B.\ S.\ Helmkamp and D.\ A.\ Browne}

\address{
Department of Physics and Astronomy\\ Louisiana State University\\
Baton Rouge LA 70803--4001}

\date{}

\maketitle

\begin{abstract}

We study how the interaction with an external incoherent environment
induces a crossover from quantum to classical behavior for a particle
whose classical motion is chaotic.  Posing the problem in the
semiclassical regime, we find that noise produced by the bath coupling
rather than dissipation is primarily responsible for the dephasing that
results in the ``classicalization'' of the particle.  We find that the
bath directly alters the phase space structures that signal the onset
of classical chaos.  This dephasing is shown to have a semiclassical
interpretation:  the noise renders the interfering paths
indistinguishable and therefore incoherent.  The noise is also shown to
contribute to the quantum inhibition of mixing by creating new
paths that interfere coherently.

\end{abstract}


\narrowtext

When a classically chaotic Hamiltonian system is treated quantum
mechanically, many of the features associated with the classical chaos
are suppressed\cite{HH}.  Since we generally expect that
classical behavior appears as the ray optics limit of quantum
mechanics, it is interesting to ask how the mixing
behavior\cite{Gutz-Reichl} of classical chaos emerges in this limit.
This question has been the subject of much recent study, leading to
arguments that coarse-graining is sufficient for classical features to
appear\cite{Casati-Paz} while others insist that the coupling to an
external environment is required\cite{Paz}.  It has also been
suggested\cite{Roncaglia} that under certain conditions the quantum
suppression of chaos will lead to a breakdown of the correspondence
principle.

In the semiclassical limit, the quantum inhibition
of mixing arises primarily from interference between classical
paths\cite{Feynman}.   Thus the proliferation of paths
associated with the onset of mixing can lead to nonclassical behavior, and
ultimately the collapse of the semiclassical picture, much sooner than
were the motion not chaotic\cite{Berry}.  Actually, this depends more on
the nature of the folding of the Lagrangian manifold in phase
space\cite{Schulman-paper}
than on a sheer abundance of paths\cite{Tomsovic}.
For example\cite{me1}, long-lived nonclassical behavior can occur
long before the chaos is fully developed
in a one-dimensional time-dependent system
when ``false caustics,'' or folds not associated with classical
turning points, appear.

In previous work\cite{me1,me2} on an isolated quantum degree of freedom,
we showed that one could identify specific quantum corrections to the
classical chaotic dynamics associated with specific features of the
classical phase space.  The quantum and classical descriptions
were essentially identical for motion in the vicinity of elliptic
fixed points, where the evolving Lagrangian manifold wraps up into
a ``whorl.''\cite{Berry}  In contrast, the quantum dynamics near the
``tendrils''\cite{Berry} associated with hyperbolic
fixed points differs strongly from the classical
description due to interference between the folds in the
Lagrangian manifold that characterize this structure.
Since these folds represent different classical
paths that reach the same final point $X$,
and the phase of each path's contribution to the semiclassical
wavefunction\cite{VV,Gutz-phs} is given by the classical action,
a nodal structure\cite{me1} appears when the difference in action between
any two folds is a smooth function of $X$
with magnitude of order $\hbar$.

Given the mechanisms that inhibit mixing in an
isolated system, one might then ask what environment-related factors
could undermine them, since we would expect\cite{Weiss} coupling
to the environment to make a quantum object behave more
classically.  For example, Caldeira and Leggett\cite{CL} showed that the
environment reduces the tunneling rate from a metastable well by
lengthening the effective distance between turning points for resonant
motion.  Presumably, such ``classicalization''\cite{Toda} by the
environment is what keeps the underlying quantum nature of macroscopic
chaos from prevailing at long times\cite{Paz,Berry}.
In this context, the questions to address are how strong must the
coupling to the environment be to accomplish this dephasing, and what
determines the characteristic dephasing time.

One way to model the environment's influence is to add a randomly
fluctuating, classical noise force to the Hamiltonian.  In this manner
Ott {\it et al.\/}\cite{Ott} found the chaotic dynamics of the quantum
kicked rotor to be strongly affected by weak amplitude-modulated noise,
especially in the semiclassical regime, while Adachi {\it et
al.\/}\cite{Toda} demonstrated that frequency-modulated noise induces
mixing behavior, even before complete classicalization occurs, for a
wide range in the noise intensity.  Scharf and Sundaram\cite{Scharf}
have shown that classical noise also destroys the
``scars''\cite{Heller-scars} (regions of enhanced amplitude that
reflect the nonmixing nature of quantum mechanics) in the
quasi-stationary\cite{Floquet} states of the quantum kicked rotor.
However, this approach does not account for the fact that the quantum
degree of freedom and the environment together form a closed system.
More properly, the environment should be treated as a large collection of
degrees of freedom that act in a deterministic fashion while reacting to
the motion of the primary degree of freedom, resulting in dissipation as
well as noise\cite{FDT}.

In this Letter we study the influence of the environment on a
quantum particle moving in a cosine potential while being driven
chaotic by an external time-dependent force.  To model the environment
more realistically, we treat
it\cite{CL,FV} as a bath of oscillators linearly coupled to the
primary degree of freedom so that when parameterized appropriately
it produces viscous damping and white noise in the classical limit.  The
hypothesis is that mixing, though disallowed in Hilbert space on first
principles, is permitted in the reduced Hilbert space of the particle
when the bath has been integrated out.

The Hamiltonian of the isolated particle is
\begin{equation}
H_o(P,X,t) = P^2 - {1\over2}\cos{(\pi X)} - \epsilon\sin{(\omega_o t)} X.
\label{H_o}
\end{equation}
We pose the problem in the naive semiclassical limit
with $\hbar = 1/200\pi$
where one can construct a well-localized wave
packet and an effectively equivalent classical probability
distribution to describe the particle's initial state.
Our initial state is a narrow Gaussian wave packet at
$(P_o, X_o)=(-0.397, 0.067)$ (see Fig.~\ref{fig-1}), and we choose
$\epsilon=0.126$ and $\omega_o=2.5$ for which the
long-time behavior\cite{LR} is chaotic.

Including the environmental degrees of freedom gives a total
Hamiltonian\cite{CL}
\begin{equation}
H=H_o +
  \sum_{\alpha} {f_\alpha\over2}\biggl\{
    { p_{\alpha}^2 \over \omega_{\alpha}^2 }
  +
  \Bigl( x_{\alpha} - X \Bigr)^2 \biggr\}.
\label{Ham_bath}
\end{equation}
In the classical limit, the particle obeys an equation of motion
with damping constant $\eta$ and noise $f(t)$
\begin{equation}
{1 \over 2} \ddot{X}=
-{ \partial {H}_o \over \partial X} -\eta \dot{X} + f(t)
\end{equation}
where the noise satisfies
\begin{equation}
\langle f(t) f(t') \rangle=2\eta k_B T\delta(t-t')\,.
\label{f_rms}
\end{equation}
To ensure this, we pick the bath frequencies $\omega_{\alpha}$
and oscillator strengths $f_{\alpha}$ to satisfy
\begin{equation}
{\pi\over 2\omega}\sum_{\alpha} f_{\alpha} \omega_{\alpha}
		    \delta (\omega_{\alpha} - \omega)
=\lim_{\omega_c\to\infty}{\eta\over 1+(\omega/\omega_c)^2}.
\label{fw_cond}
\end{equation}
We characterize the coupling strength to the bath via the
$Q=\Omega_o/2\eta$ of the oscillator, where $\Omega_o=\pi$ is the
frequency of small oscillations in the bottom of the well, and
concentrate here on the weak-coupling limit $Q\gg1$.

Since our $H_o$ has a large time-dependent term, neither an
imaginary-time path integral formulation of the reduced density
matrix\cite{CL} nor linear response techniques for treating
time-dependent many-body Hamiltonians at finite
temperatures\cite{Weiss} are suitable.  We therefore
studied\cite{mythesis} this problem numerically
using a large number of classical oscillators ($N$=10~000)
with fixed oscillator strengths and randomly distributed frequencies
chosen to agree with the distribution of Eq.~(\ref{fw_cond}), which
is justified in the semiclassical limit where the bath is essentially
a collection of almost-coherent states\cite{LLQM}, each following
the classical equations of a very weakly perturbed harmonic oscillator.
As the collision rate of the particle with the bath,
$\omega_c$ in Eq.~(\ref{fw_cond}) is chosen to be
large compared to the natural frequency of $H_o$ with
$\omega_c=10\Omega_o$.

Finite $N$ means that a given realization of the dynamics will be
initial condition-dependent, since all possible realizations of initial
conditions for the bath are not represented with a finite number of
oscillators.  Thus one must finish ``tracing out'' the initial
conditions by averaging over runs, though in practice the need for this
depends on the amount of noise in the system as well as $N$.
(With $N=10~000$, about 25 runs were needed before our results
stabilized for $Q=10^3$, as compared to a single run for $Q \geq 10^4$.)
Furthermore, because the mean energy per oscillator is not vanishingly
small, having $N$ finite sets a minimum temperature---with
$\langle H_o \rangle$ a few $k_BT$ above the ground state---so that the bath
on average takes energy from the particle.  Finally, by requiring that
$\omega_c$ be no larger than $k_BT/\hbar$, we ensure that even the
highest frequency oscillators are relatively classical, but with $\omega_c
> \Omega_{o}$ by an order of magnitude, this also
means the temperature is high on the natural scale of the problem.

We use a simple one-step predictor-corrector
algorithm\cite{mythesis} to evolve the combined system of quantum
particle and semiclassical oscillators forward in
time.  For the corresponding classical motion we use the
phenomenological equations of motion to integrate forward a cloud of
initial conditions that represents the phase space probability density
$\rho_{cl}(P,X,t)$.  This two-dimensional distribution is depicted
graphically as a dot plot at snapshots in time and also in reduced form
$\rho_{cl}(X,t)=\int\rho_{cl}(P',X,t){\rm d}P'$ as a one-dimensional
histogram.  The ``wave function'' $\psi(X,t)$ is then compared to
$\rho_{cl}(P,X,t)$ in two ways: by resampling the quantum amplitude
$|\psi(X,t)|^2$ to match the bin size of the classical histogram and by
projecting $\psi(X,t)$ into phase space using the Husimi
transform\cite{Husimi}.

In Fig.~\ref{fig-2} we compare the one-dimensional distributions
$|\psi(X)|^2$ and $\rho_{cl}(X)$ for $Q=\infty$, $Q=10^4$, and
$Q=10^3$ at a time $t=13$.   The initially equivalent quantum and
classical descriptions have begun to differ because of the appearance
of the first tendril, which is the folded feature in
Fig.~\ref{fig-2}(a) near $P= -1.0$ superimposed on the spiraling
whorl in the region $0.1 < X < 0.4$.	The resulting nonclassical nodal
structure represents a semiclassical beating
phenomenon between the upper branch of the tendril and the remnant separatrix,
as discussed earlier.

Note that these nodes, though largely unaffected by the environment for
$Q \geq 10^4$ (Fig.~\ref{fig-2}(b)), seem to have disappeared
altogether for $Q=10^3$ (Fig.~\ref{fig-2}(c)).  This result is
interesting, not so much because the presence of an environment
suppresses quantum interference (which one expects), but because it
happens barely six cycles into the motion, long before dissipation is
an appreciable effect ($t \ll Q/\Omega_o$).  This is
particularly apparent when comparing Figs.~\ref{fig-2}(a)
and \ref{fig-2}(c): the nodes disappear
without a noticeable shift in the manifold, indicating that
the energy of each classical trajectory does
not change appreciably. Thus we infer
that noise rather than dissipation plays the dominant role in the
semiclassical regime for weak damping, where
the noise strength varies as $Q^{-1/2}$ while
the dephasing by energy loss should vary as $Q^{-1}$ for large $Q$.

If this semiclassical interpretation of the
dephasing is correct, one would expect to see nodal structure associated
with the tail of the tendril ($X > 0.4$) where the paths do resolve and
the action difference is presumably smooth enough to preserve the
coherence.  While obscured by a ``turning point'' in  Fig.~\ref{fig-2}(c),
these nodes do in fact appear in the Husimi transform
shown in Fig.~\ref{fig-3}.  Note in particular that the primary antinode
at $(P,X) \approx (-0.75,0.45)$ in Fig.~\ref{fig-3}(a) coincides
with where the noisy paths coalesce in Fig.~\ref{fig-3}(b)
rather than at the leading fold of the tendril ({\em i.e.}, the
false caustic) where it occurs for the isolated problem.  In addition,
we calculated the action difference and found it to be a
well-defined quantity in the region where these nodes are observed
with the enclosed phase space correctly accounting for the number
of nodes\cite{me1}.

The surprising feature of this structure is that the paths that interfere to
produce it would not exist at all without the noise; that is, evolving the
cloud with damping but without noise produces no appreciable amplitude
here, only the usual (nominal) exponential tail.
Moreover, this amplitude does not come from a specific region of
the cloud.  Rather, cloud particles are kicked (or ``filtered'') here
``at random'' while in the vicinity of the hyperbolic fixed point.

In summary, when we introduce the coupling to the environment, we find
that quantum coherence is destroyed for the tendrils first, and that this
dephasing is primarily a noise effect.  All paths are not affected
equivalently, however: some paths retain their coherence, including
certain ``new'' paths that appear because of the noise, while others do
not.  Noise in ``quantum chaos,'' like classical
chaos\cite{noise-mixing}, thus has a dual role: it assists the mixing
by scrambling existing paths but also inhibits the
mixing by creating new paths that interfere. These new paths, having crossed
the (dynamic) potential barrier associated with a classical turning
point, represent noise-assisted tunneling. It is interesting that
while these paths arise from the coupling to the environment, they are not
dephased as much as those paths that are present without the coupling.

This work was supported by the National Science Foundation under
Grant No.~DMR--9408634.

\begin{figure}
\caption{
(a) The unperturbed cosine potential with a schematic
depiction of the initial wave packet---a coherent state ``centered''
at $(P_o, X_o)=(-0.397, 0.067)$ with width corresponding to the
ground-state wave function.
(b) The initial orbit at $E=H_o(P_o,X_o)=-0.331$ (dashed) and the
unperturbed separatrix at $E=0.500$ (solid).  The size and ``location''
of the initial state are given by the three-sigma ellipse at $(P_o, X_o)$.
Note the locations of the hyperbolic fixed points (denoted {\em hfp}).
}
\label{fig-1}
\end{figure}

\begin{figure}
\caption{
A comparison of $|\psi(X,t)|^2$ (solid) and $\rho_{cl}(X,t)$ (dashed)
at $t=13$ for (a) $Q=\infty$, (b) $Q=10^4$, and (c) $Q=10^3$.  In each
case the classical dot plot is shown as well.  Note that the oscillations
in $|\psi |^2$ corresponding to the folded tendril
feature disappear when $Q$ is increased from $10^3$ to $10^4$ (roughly
a three-fold increase in root-mean-square noise).
}
\label{fig-2}
\end{figure}

\begin{figure}
\caption{
The (a) Husimi transform and (b) classical dot plot corresponding
to Fig.~\protect\ref{fig-2}(c) magnified in the region of the tendril.
(Contours are on a logarithmic scale.) Note the absence of structure
in the Husimi transform at the old location of the false caustic.  Also
note that the primary antinode occurs where the muddled paths resolve.
}
\label{fig-3}
\end{figure}

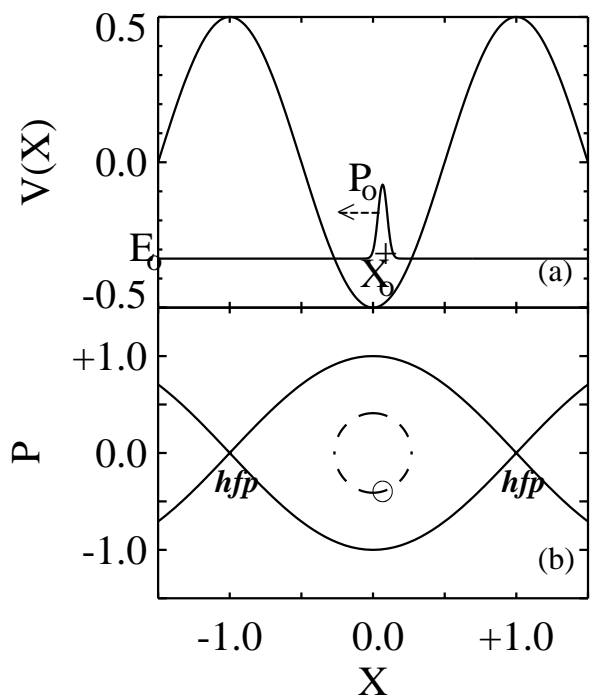
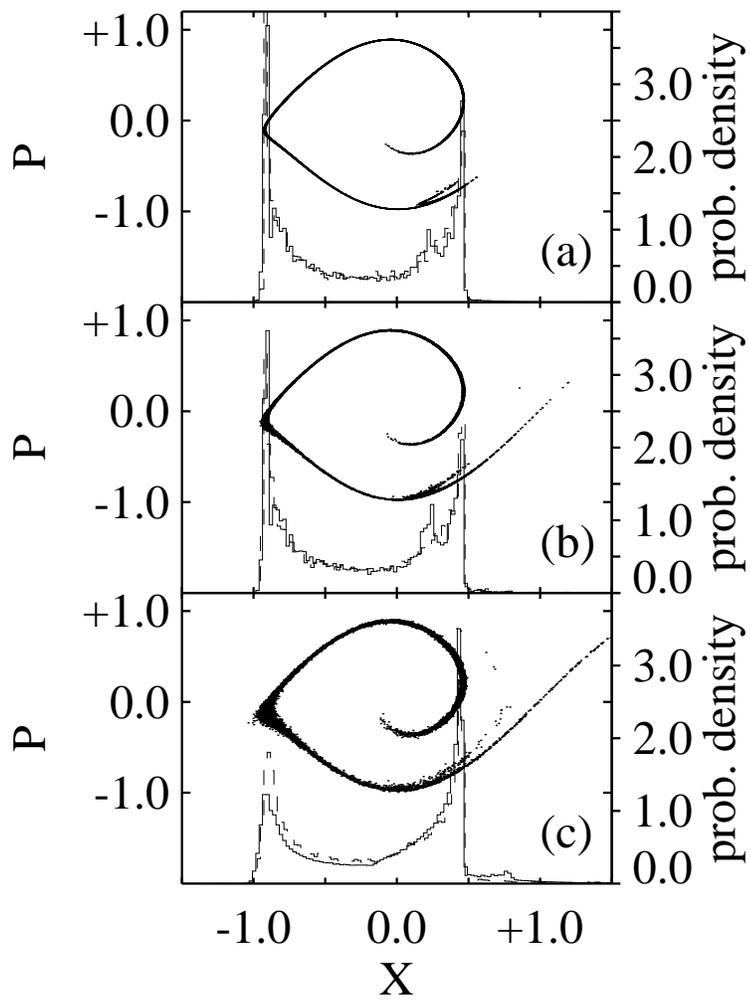
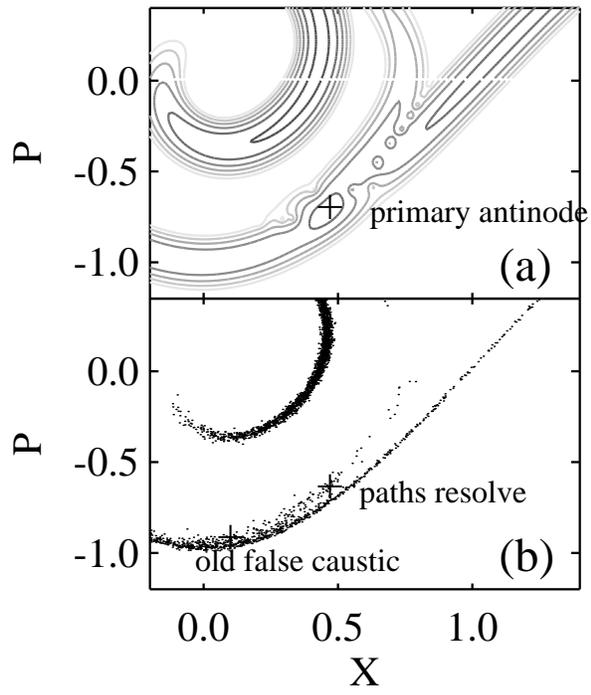
\end{document}